# Brief review on iron-based superconductors: are there clues for unconventional superconductivity?

Hyungju Oh[a], Jisoo Moon[a], Donghan Shin[a], Chang-Youn Moon[b], Hyoung Joon Choi[*,a]

[a] *Department of Physics and IPAP, Yonsei University, Seoul 120-749, Korea*
[b] *Department of Chemistry, Pohang University of Science and Technology, Pohang, 790-784 Korea*

December 30, 2011

**Abstract**

Study of superconductivity in layered iron-based materials was initiated in 2006 by Hosono's group, and boosted in 2008 by the superconducting transition temperature, $T_c$, of 26 K in LaFeAsO$_{1-x}$F$_x$. Since then, enormous researches have been done on the materials, with $T_c$ reaching as high as 55 K. Here, we review briefly experimental and theoretical results on atomic and electronic structures and magnetic and superconducting properties of FeAs-based superconductors and related compounds. We seek for clues for unconventional superconductivity in the materials.

## I. Introduction

Since the discovery of superconductivity in mercury, it has been a key interest to find the origin of the superconductivity and materials with higher superconducting transition temperature ($T_c$). Up to now, $T_c$ is the highest in copper-oxide perovskite materials [1] with values higher than 150 K achieved in 1980s and 1990s, but the origin of the high-$T_c$ superconductivity in the copper oxides is still not understood. After that, superconductivity in MgB$_2$ ($T_c$ = 39 K) found in 2001 renewed interest in maximal $T_c$ by conventional phonon-mediated superconductivity, and more recently, discovery of superconductivity in iron pnictides [2, 3] opened a new family of unconventional superconductors having $T_c$ relatively higher than conventional superconductors. Iron-pnictide superconductors have layer structures of FeAs or FeP, and magnetism from 3$d$ electrons is closely related with superconductivity.

## II. FeAs-based and related materials

Iron-based superconductors started with the discovery of superconductivity at 4 K in LaFePO in 2006 [2], and great interests have been drawn since 2008 when $T_c$ was raised to 26 K in LaFeAsO$_{1-x}$F$_x$ by replacing phosphorous with arsenic, and some of oxygen with fluorine [3]. So far, iron-based superconductors have been extended to a large variety of materials including four prototypical families of iron-based superconductors, 1111, 122, 111, and 11 types, as shown in Fig. 1, and further variations such as 42622-type iron pnictides [4-7] and 122-type iron chalcogenides [8-11].

*1111-type family.*—The 1111-type family includes LaFePO and LaFeAsO$_{1-x}$F$_x$, which are mentioned above, and LnFeAsO with various lanthanide elements (Ln). The atomic structure of the 1111 family consists of negatively charged FeP or FeAs layers, where Fe atoms form a planar square lattice, and positively charged LnO layers, as shown in Fig. 1(a). With or without doping, electrons are conducting in FeP or FeAs layers. Figs. 2(a) and (b) show that the electrical resistivity of pure LaFePO drops at 4 K and that of F-doped LaFePO drops at higher temperature. These superconducting transitions are confirmed by magnetic susceptibility [Fig. 2(c)]. It is noticeable that decrease of the resistivity starts at ~ 10 K in F-doped LaFePO. Unlike LaFePO, undoped LaFeAsO does not show superconductivity [Fig. 3(a)]. With doping of F replacing O in part, $LaFeAsO_{0.89}F_{0.11}$ becomes superconducting. When small pressure is applied to $LaFeAsO_{0.89}F_{0.11}$, $T_c$ increases, reaching a maximum value of $T_c$ = 43 K at 4 GPa, and then it decreases to $T_c$ = 9 K at 30 GPa [12]. At ambient pressure, $T_c$ higher than 40 K is achieved in $SmFeAsO_{1-x}F_x$ [13].

---

*Corresponding author.   Fax : +82 2 392 1592
e-mail : h.j.choi@yonsei.ac.kr

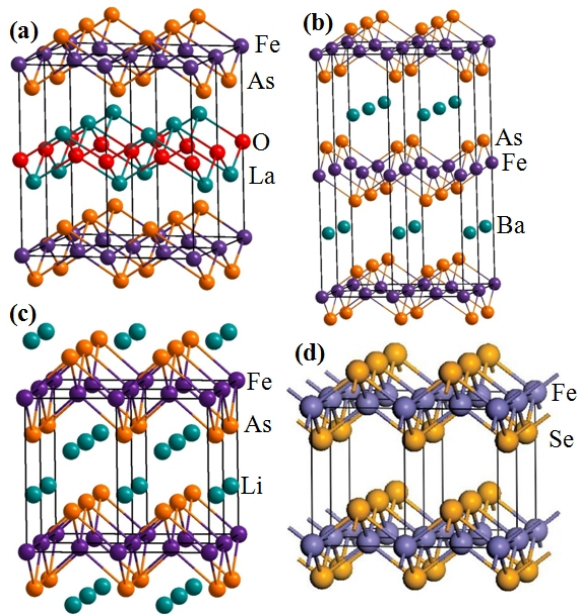

Fig. 1. Four families of iron-based superconductors, (a) 1111, (b) 122, (c) 111, (d) 11 type. Fig. (d) reprinted from Ref. 22: F.-C. Hsu *et al.*, Proc. Natl. Acad. Sci. U.S.A. **105**, 14262 (2008). Copyright 2008 by the National Academy of Sciences.

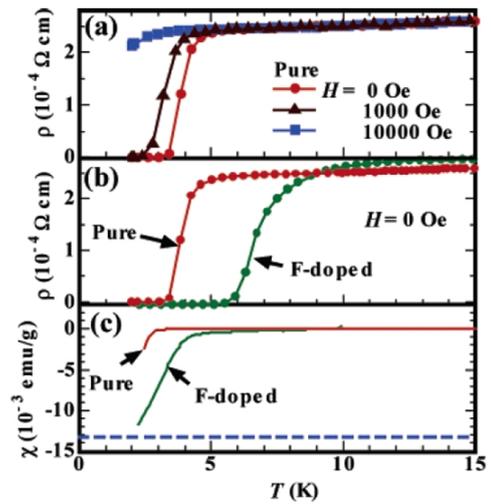

Fig. 2. (a) Electrical resistivity, $\rho$, vs. temperature, $T$, for pure LaFePO at various magnetic field, $H$. (b) $\rho$ vs $T$ and (c) magnetic susceptibility, $\chi$, vs $T$ for pure and F-doped LaFePO. Fig. reprinted from Ref. 2: Y. Kamihara *et al.*, J. Am. Chem. Soc. **128**, 10012 (2006). Copyright 2006 by the American Chemical Society.

Reported values of $T_c$ in 1111-type materials include 4 K in LaFePO [2], 26 K in $LaFeAsO_{0.89}F_{0.11}$ [3], 41 K in $CeFeAsO_{0.84}F_{0.16}$ [14], 52 K in $PrFeAsO_{0.89}F_{0.11}$ [15], 54.3 K in $NdFeAsO_{1-y}$ [16], 55 K in $SmFeAsO_{0.9}F_{0.1}$ [17], and 54 K in $GdFeAsO_{1-y}$ [16].

*122- and 111-type families.*—Right after discovery of 1111-type family, $Ba_{1-x}K_xFe_2As_2$ (122 type) with $T_c$ of 38 K [18] were found, followed by LiFeAs (111 type) with $T_c$ of 18 K [19]. The 122- and 111-type families have simpler structures than the 1111-type family. While FeAs or FeP layers are present in 1111-, 122-, and 111-type materials, the 'blocking layer' which separates FeAs or FeP is different for each type: rare-earth oxide or fluoride (for example, LaO or SrF) for the 1111-type family, alkaline-earth metals (for example, Ba) for the 122-type family, and alkali metals (for example, Li) for the 111-type family. Reported values of $T_c$ in 122-type materials include 38 K in $Ba_{0.6}K_{0.4}Fe_2As_2$ [18], 32 K in $Sr_{0.6}K_{0.4}Fe_2As_2$ [20], 26 K in $Sr_{0.6}Na_{0.4}Fe_2As_2$ [21], and 21 K in $Ca_{0.6}Na_{0.4}Fe_2As_2$ [20].

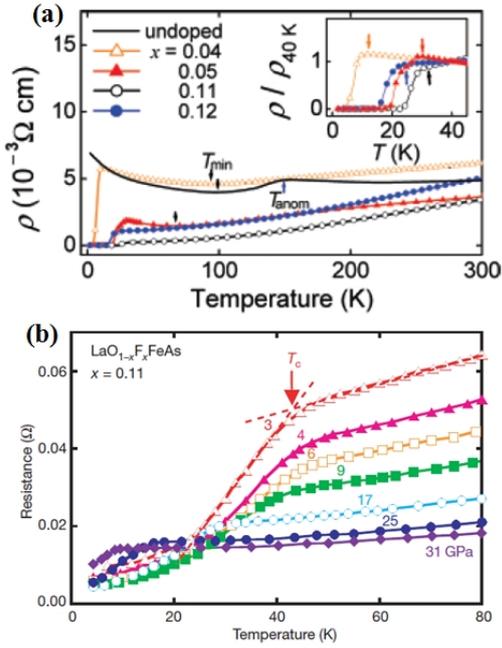

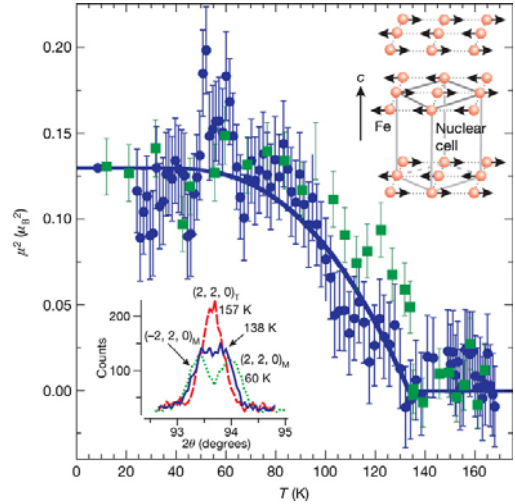

Fig. 3. (a) Electrical resistivity, $\rho$, vs temperature in LaFeAsO$_{1-x}$F$_x$ for x = 0.0, 0.04, 0.05, 0.11, and 0.12. Fig. reprinted from Ref. 3: Y. Kamihara *et al.*, J. Am. Chem. Soc. **130**, 3296 (2008). Copyright 2008 by the American Chemical Society. (b) Temperature dependence of electrical resistance above 3 GPa in LaFeAsO$_{0.89}$F$_{0.11}$. The maximum $T_c$ is 43 K at 4 GPa. Fig. reprinted from Ref. 12: H. Takahashi *et al.*, Nature **453**, 376 (2008). Copyright 2008 by Macmillan Publishers Ltd.

*11-type family.*—The 11-type materials are iron chalcogenide, which started with FeSe having $T_c$ of 8 K at ambient pressure [22] and 36.7 K with applied pressure of 8.9 GPa [23]. This family also includes FeTe$_{1-x}$Se$_x$, and FeTe$_{1-x}$S$_x$. These materials have the simplest structure among iron-based superconductors, in which iron-chalcogenide layers are simply stacked together. Reported values of $T_c$ in 11-type materials include 8 K in FeSe at ambient pressure [22] and 36.7 K in FeSe under pressure of 8.9 GPa [23], as mentioned above, and 14 K in FeTe$_{0.5}$Se$_{0.5}$ [24], 2 K in Fe$_{1.13}$Te$_{0.85}$S$_{0.1}$ [25], and 10 K in FeTe$_{0.8}$S$_{0.2}$ [26].

Fig. 4. Antiferromagnetic ordering in LaFeAsO. The solid line is a simple fit to mean field theory that gives a Néel temperature $T_N$ = 137 K. The top-right inset shows the single-stripe-type antiferromagnetic ordering of Fe magnetic moments. Fig. reprinted from Ref. 27: C. de la Cruz *et al.*, Nature **453**, 899 (2008). Copyright 2008 by Macmillan Publishers Ltd.

### III. Phase diagram

*Different behaviors of undoped compounds.*—The undoped parent compounds of iron-based superconductors are either superconducting or antiferromagnetic at low temperatures. LaFePO, LiFeAs, and FeSe, for example, are nonmagnetic and superconducting even without doping. In contrast, undoped LaFeAsO and BaFe$_2$As$_2$, for example, are non-superconducting antiferromagnetic metals, and electron or hole doping suppresses the magnetic order and induces superconductivity.

*Structural and magnetic transitions in 1111-type family.*—The undoped 1111-type iron pnictides show structural and magnetic phase transitions at slightly different temperatures. It is reported by neutron scattering experiments [27] that LaFeAsO undergoes an abrupt structural transition at 155 K, changing from high-temperature tetragonal structure (space group P4/nmm) to low-temperature monoclinic structure (space group P112/n). At ~137 K, long-range antiferromagnetic order starts to develop with a small moment of 0.36 $\mu_B$ per atom and a simple antiferromagnetic ordering of single-stripe type [27], as shown in Fig. 4. The direction of the Fe magnetic moment was later measured to be longitudinal in the sense that it is parallel or anti-parallel to the antiferromagnetic ordering wave vector [28]. With electron or hole doping, the antiferromagnetic order in the parent compounds is suppressed and superconductivity emerges. Fig. 5 shows the typical phase diagram for the 1111-type family which is determined by neutron-scattering measurements on $CeFeAsO_{1-x}F_x$ [29].

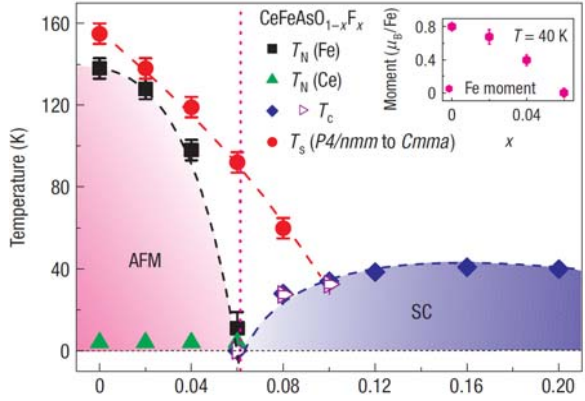

Fig. 5. Phase diagram of $CeFeAsO_{1-x}F_x$. The red circles indicate the onset temperature of structural transition. Antiferromagnetic and superconducting phases are marked with AFM and SC, respectively. Fig. reprinted from Ref. 29: J. Zhao *et al.*, Nature Mater. **7**, 953 (2008). Copyright 2008 by Macmillan Publishers Ltd.

*Typical phase diagram for 122-type family.*─A compilation of experimental phase diagrams [30-33] is presented in Fig. 6(a) for the Ba-based 122 system, which captures the main traits of 122-type family. In $BaFe_2As_2$, the systematic replacement of either the alkaline-earth metal (Ba), transition metal (Fe), or pnictogen (As) with different elements almost universally suppresses the non-superconducting antiferromagnetic state of parent compounds to a superconducting nonmagnetic state.

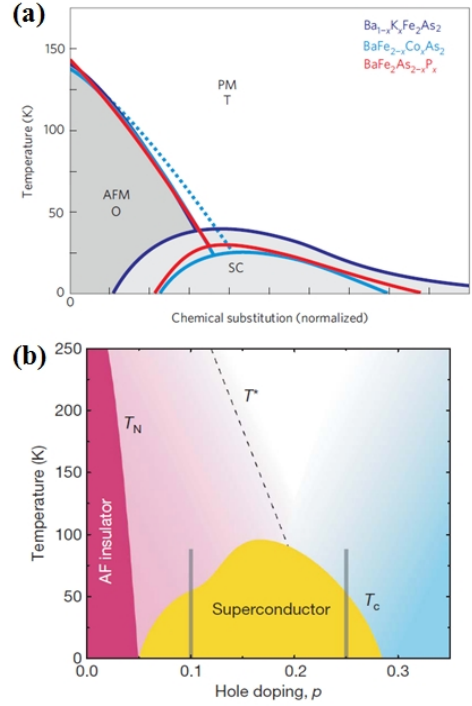

Fig. 6. (a) Phase diagram of the $BaFe_2As_2$ system, shown for K [30], Co [31] and P [32] doping. The dotted line indicates tetragonal (T) to orthorhombic (O) structural phase transitions in Co doped samples [31]. Fig. reprinted from Ref. 33: J. Paglione and R. L. Greene, Nature Phys. **6**, 645 (2010). Copyright 2010 by Macmillan Publishers Ltd. (b) Doping dependent antiferromagnetic transition temperature, $T_N$, superconducting transition temperature, $T_c$, and pseudogap crossover temperature, $T^*$, in YBCO [35]. Fig. reprinted from Ref. 35: N. Doiron-Leyraud *et al.*, Nature **447**, 565 (2007). Copyright 2007 by Macmillan Publishers Ltd.

This phase transition from the antiferromagnetic state to the superconducting state is generic property of the iron-based superconductor systems which can also be produced by applied external pressure [34]. It is remarkable that the phase diagram of the iron-based superconductor system is very similar to that of copper-oxide materials [35] shown in Fig. 6(b), except that the undoped parent compound of copper-oxide materials is an antiferromagnetic Mott

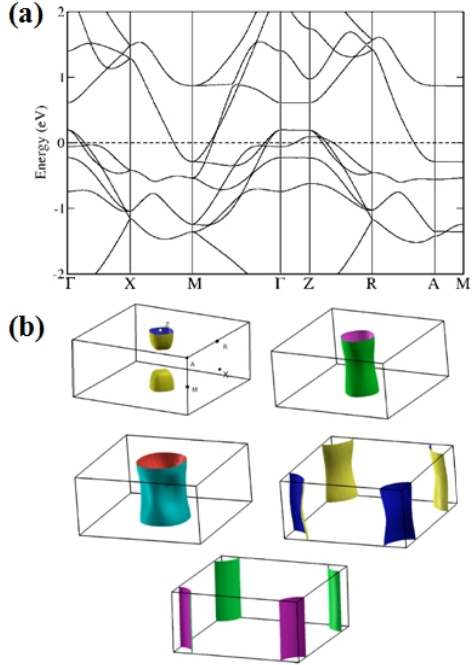

Fig. 7. (a) Calculated band structure and (b) the Fermi surface of LaFePO from DFT. Fig. reprinted from Ref. 36: S. Lebègue, Phys. Rev. B **75**, 035110 (2007). Copyright 2007 by the American Physical Society.

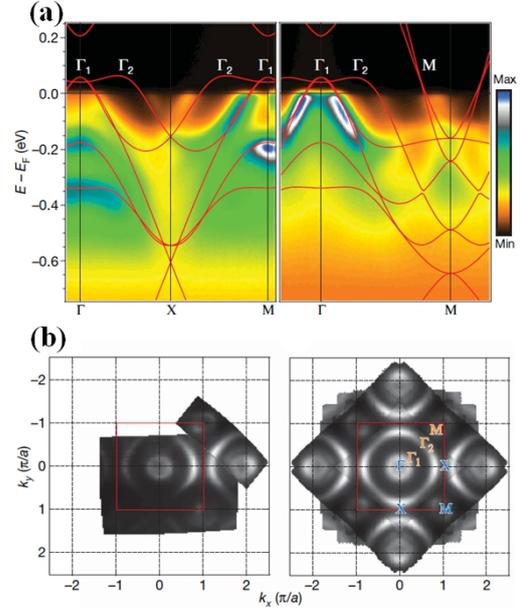

Fig. 8. (a) ARPES band dispersion and (b) unsymmetrized (left) and symmetrized (right) Fermi surface maps of LaFePO. Fig. reprinted from Ref. 37: D. H. Lu *et al.*, Nature **455**, 81 (2008). Copyright 2008 by Macmillan Publishers Ltd.

insulator.

## IV. Electronic structures

*DFT study of LaFePO.*—The electronic structure of iron-based materials was first studied by S. Lebègue in 2007 using *ab initio* calculations based on the density functional theory (DFT) [36]. The calculated band structure and the Fermi surface of LaFePO are shown in Figs. 7(a) and (b), respectively. The Fermi surface of LaFePO consists of five sheets, resulting from five bands which cross the Fermi level. Four of the five sheets are cylinder-like along the $k_z$ direction reflecting quasi-two-dimensional nature of the atomic structure: two of them are of hole type along the Γ-Z high-symmetry line of the Brillouin zone and the other two are of electron type along the M-A line. The 5th sheet is a distorted hole-type sphere centered at the Z high-symmetry point.

*ARPES band dispersion of LaFePO.*—In 2008 after the discovery of superconductivity in LaFeAsO$_{1-x}$F$_x$, angle-resolved photoemission spectroscopy (ARPES) of LaFePO was performed [37] and compared with the calculated band structure. As shown in Fig. 8(a), it turns out that the ARPES band dispersions are similar to the DFT band dispersions but the ARPES bandwidths near the Fermi level are quite smaller than DFT results. Observed ARPES band dispersions and Fermi-surface topology (Fig. 8) are consistent with the DFT prediction [36] after shifting the DFT bands up by 0.11 eV and then shrinking the DFT bandwidths by a factor of 2.2 [37]. The overall level of agreement between the experiments and the adjusted calculational results is significant, indicating that the DFT framework, in which electrons are itinerant, captures the essence of electronic states

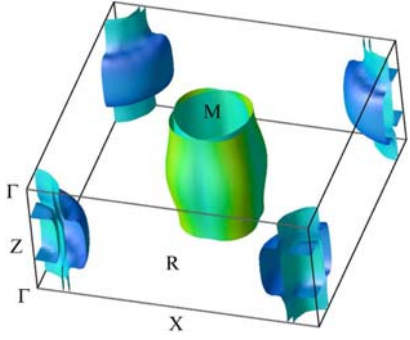

Fig. 9. Fermi surface LaFeAsO in non-magnetic phase. Fig. reprinted from Ref. 38: D. J. Singh and M.-H. Du, Phys. Rev. Lett. **100**, 237003 (2008). Copyright 2008 by the American Physical Society.

near the Fermi level in LaFePO.

*DFT and AIPES of LaFeAsO.*—The electronic structure of LaFeAsO calculated with DFT [38] is similar to that of LaFePO [36], as shown in Fig. 9. High-resolution ARPES of LaFeAsO is difficult to obtain for lack of high-quality single crystals of LaFeAsO. Instead, Fig. 10 shows comparison of angle-integrated photoemission spectra (AIPES) with the density of states obtained from DFT for LaFePO and LaFeAsO [39]. Both AIPES results consist of a sharp intense peak near the Fermi level that is separated from the main valence band peaks at higher binding energy. According to the DFT calculations, states near the Fermi level have dominantly Fe $d$ characters while the peaks at higher binding energy are mixtures of O $p$ states and hybridized Fe $d$ and pnictogen $p$ states. Compared with the calculated density of states, the measured peak near the Fermi level in each material has a narrower width than the calculated Fe $d$ states and it is pushed closer to the Fermi level, which is consistent with the band renormalization effect mentioned above for ARPES of LaFePO. The valence-band peaks at higher binding energy, on the other hand, are shifted towards higher binding energy, resulting in slightly larger total valence-band width.

*DFT+DMFT studies.*—The combination of density functional theory and dynamical mean-field theory (DFT+DMFT) [40] is used to study the electronic structures in non-magnetic LaFePO [41]. Calculated momentum-resolved spectral functions and effective electron mass enhancement are in good quantitative agreement with corresponding experimental data, showing that the bandwidth renormalization near the Fermi level may originate from the dynamical correlation effects although the renormalization may not be intrinsically connected with the value of $T_c$ [41].

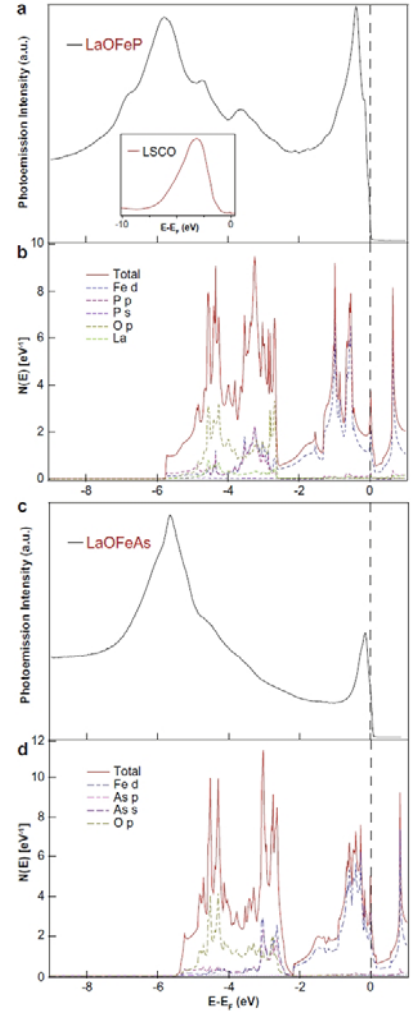

Fig. 10. (a) Angle-integrated photoemission spectrum (AIPES) and (b) calculated density of states, $N(E)$, of LaFePO. (c) AIPES and (d) $N(E)$ of LaFeAsO. The inset in (a) shows the valence band of LSCO for comparison. Fig. reprinted from Ref. 39: D. H. Lu *et al.*, Physica C **469**, 452 (2009). Copyright 2009 by Elsevier B.V.

## V. Magnetic ordering

For the origin of the magnetism in iron-based superconductors, itinerant-electron magnetism and local-moment magnetism are widely studied.

*Itinerant-electron magnetism.*─Magnetism from itinerant electrons was proposed by theories [42-46] and experiments [27, 46-50], because the cylinder-like electron-type pockets of the Fermi surface and hole-type ones are separated by a two-dimensional commensurate nesting vector, $(\pi, \pi)$, in the Brillouin zone, as shown in Fig. 9, and this nesting vector is the same as commensurate antiferromagnetic ordering wave vector except for some 11-type iron-based superconductors [51, 52]. Furthermore, the itinerant picture is supported by the reduced magnetic moments and the decrease of the density of states near Fermi level in the magnetic phase of iron pnictides.

*Local-moment magnetism.*─Magnetism from local moments is based on the Heisenberg-type interaction between localized spin moments [53-56]. In this picture, the observed antiferromagnetic ordering of single-stripe type in iron pnictides is a frustrated spin configuration which occurs when the exchange interaction ($J_2$) between the next nearest neighbors is larger than half of the nearest-neighbor exchange interaction ($J_1$).

*Combined models.*─Magnetism in iron-based superconductors may have both itinerant-electron and local-moment characters. In this point of view, many theories and models have been proposed in which itinerant electrons and localized moments coexist in iron-based superconductors and they play some roles in magnetism [57-63]. Dual nature of local moments and itinerant electrons in magnetism is also proposed using x-ray emission spectroscopy [64].

*Magnetism in iron chalcogenides.*─In the cases of FeTe and FeSe, the calculated electronic band structures in nonmagnetic phase [65], especially the Fermi surfaces, are similar to those of LaFeAsO and $BaFe_2As_2$. This expects that FeTe and FeSe would have the single-stripe-type antiferromagnetic order if the magnetism is induced itinerantly by the Fermi surface nesting. However, it was predicted by DFT calculations [52] and confirmed by experiments [51] that the magnetic ground state of FeTe has a double-stripe-type antiferromagnetic order in which the magnetic moments are aligned ferromagnetically along a diagonal direction and antiferromagnetically along the other diagonal direction of the Fe square lattice, as shown schematically in Fig. 11(a). Since FeTe also has the $(\pi, \pi)$ Fermi surface nesting similar to other iron-based superconductors [65], the observed double-stripe-type magnetic ordering in FeTe, which corresponds to a wavevector of $(\pi, 0)$, may indicate irrelevance of Fermi-surface nesting mechanism at least for FeTe [52]. A recent first-principles study [66] suggested the emergence of $(\pi, 0)$ Fermi-surface nesting from the assumption of substantial electron doping by excess Fe; however, this doping effect was not observed in ARPES experiments [67, 68]. Meanwhile, DFT calculations predict that the ground state of FeSe has the single-stripe-type antiferromagnetic order, similar to those in LaFeAsO and $BaFe_2As_2$, as shown in Fig.

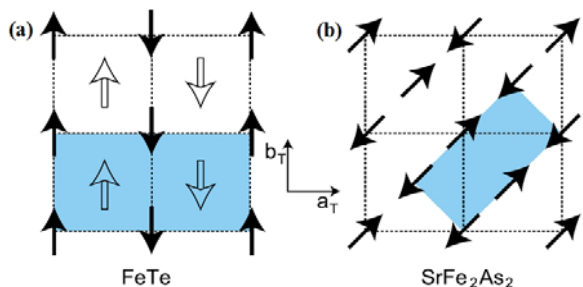

Fig. 11. (a) Double-stripe-type antiferromagnetic order in FeTe. The solid and hollow arrows represent two sublattices of spins. (b) Single-stripe-type antiferromagnetic order in $SrFe_2As_2$. The shaded area indicates the magnetic unit cell. Fig. reprinted from Ref. 51: S. Li *et al.*, Phys. Rev. B **79**, 054503 (2009). Copyright 2008 by the American Physical Society.

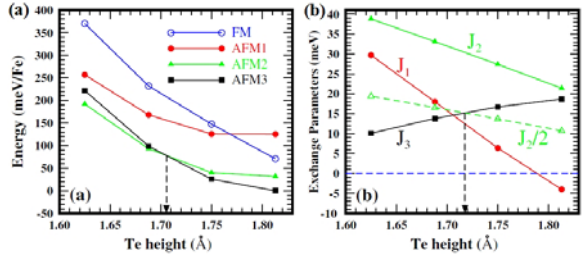
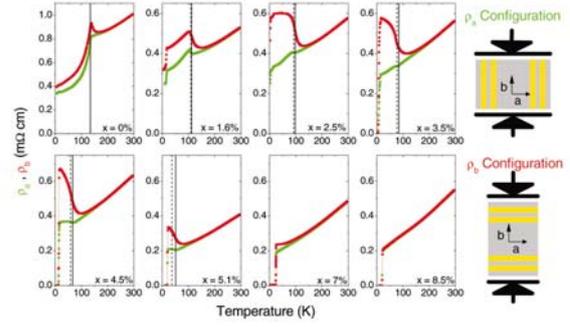

Fig. 12. Magnetic order switching in FeTe by Te height. (a) Total energy vs Te height calculated for ferromagnetic ordering (FM) and checker-board, single-stripe, and double-stripe antiferromagnetic orderings (AFM1, AFM2, AFM3, respectively). (b) Exchange parameters $J_s$ vs Te height. Vertical arrows denote the critical Te height where the AFM2-AFM3 switch takes place. Fig. reprinted from Ref. 69: C.-Y. Moon and H. J. Choi, Phys. Rev. Lett. **104**, 057003 (2010). Copyright 2010 by the American Physical Society.

11(b), although FeSe does not show any magnetic ordering in experiments.

*antiferromagnetic order switching in iron chalcogenides.*—The energetic stability of $(\pi, 0)$ antiferromagnetic ordering over $(\pi, \pi)$ ordering in FeTe can be effectively described by the nearest, second nearest, and third nearest neighbor exchange parameters, $J_1$, $J_2$, and $J_3$, respectively, with the condition $J_3 > J_2/2$ [52]. C.-Y. Moon and H. J. Choi [69] found that Te height from the Fe plane is a key factor that determines antiferromagnetic ordering patterns in FeTe [69], so that the ground-state magnetic ordering changes from the $(\pi, 0)$ with the optimized Te height to the $(\pi, \pi)$ patterns when Te height is lowered [Fig. 12(a)]. Fig. 12(b) shows the three exchange parameters $J_1$, $J_2$, and $J_3$, as functions of Te height. As Te height decreases, $J_1$ and $J_2$ increase while $J_3$ decreases, and when Te height is 1.72 Å, $J_2/2$ starts to exceed $J_3$ which reflects that the $(\pi, \pi)$ antiferromagnetic ordering becomes more stable than the $(\pi, 0)$. As the Te atom gets close to the Fe plane, the Fe-Te-Fe angle approaches 180°, maximizing the overlap between Te 5$p$ and Fe 3$d$ orbital lobes and thereby increasing $J_1$ and $J_2$ which have superexchange nature [53]. As for $J_3$, a longer range interaction of RKKY type is needed in which local spin moments can be coupled via itinerant electrons near the Fermi level. When Te is lowered, the density of states at the Fermi level is greatly reduced [69], which is consistent with the decrease of $J_3$, for the number of itinerant electrons mediating long-range local-moment interaction would decrease as the density of states at the Fermi level reduces.

Fig. 13. Temperature dependence of the in-plane resistivity $\rho_a$ (green) and $\rho_b$ (red) of detwinned Ba(Fe$_{1-x}$Co$_x$)$_2$As$_2$ for $x = 0$ to 0.085. Solid and dashed vertical lines mark critical temperatures for the structural and magnetic phase transitions $T_S$ and $T_N$, respectively. Diagrams on the right illustrate experimental setup for four-probe measurement under uniaxial compression. Fig. reprinted from Ref. 83: J.-H. Chu *et al.*, Science **329**, 824 (2010). Copyright 2010 by American Association for the Advancement of Science.

## VI. In-plane anisotropy in antiferromagnetic phase

The antiferromagnetic transition in iron-based superconductors is always preceded by or coincident with a tetragonal to orthorhombic structural transition as the temperature is lowered [27, 29, 70-72]. It has been proposed that this structural transition is driven by an electronic phase transition [73-76], perhaps due to orbital ordering [77-80] or fluctuating antiferromagnetism [73, 74]. In both proposals, a large in-plane electronic anisotropy was anticipated. In the orthorhombic (and magnetic) phase of iron-based superconductors, the system forms twinned crystal and magnetic domains, with the axes for the two domains orthogonal to each other [81]. The existence of such twin domains is not a problem for nanoscale probes such as the scanning tunneling microscope [82], but detwinning is needed for transport measurements [83] and ARPES [84-86] because informations from the two domains would be mixed.

*Resistivity of detwinned samples.*—Fig. 13 shows resistivity of $Ba(Fe_{1-x}Co_x)_2As_2$ as functions of temperature for $x$ ranging from the undoped parent compound ($x = 0$) to fully overdoped composition ($x = 0.085$) [83]. To detwin the sample in the antiferromagnetic phase, stress is applied along an edge of the square lattice of Fe atoms (i.e., along one of the two nearest-neighbor directions in the iron square lattice), reducing the $C_4$ symmetry of the iron lattice to $C_2$. The longer $a$ axis of the rectangular unit cell is perpendicular to the stress while the shorter $b$ axis is parallel to the stress. At low temperature, Fe magnetic moments are ordered antiferromagnetically along the longer $a$ axis and ferromagnetically along the shorter $b$ axis. The resistivity measurement of detwinned samples reveals that the resistivity along the longer $a$ axis, $\rho_a$, is smaller than $\rho_b$ ($\rho_a < \rho_b$) for all underdoped compositions [83]. This is somewhat counterintuitive, since the smaller orbital overlap and the antiferromagnetic ordering along the longer $a$ axis might produce more resistance.

*ARPES of detwinned samples.*—ARPES experiments are performed on mechanically detwinned $BaFe_2As_2$ and $Ba(Fe_{1-x}Co_x)_2As_2$ [85, 86]. Fig. 14(a) is the Fermi surface map for a detwinned sample [86]. The band dispersions along the Γ-X in Fig. 14(b) are not significantly different from those obtained from twin-domain samples [87]. Most notably, split bands at the X point that appear below the magnetic transition still exist after detwinning, indicating that they are not from different domains but they are genuine features of a magnetically ordered state inside a domain. In contrast, away from the Γ-X line, the band dispersions along lines parallel to the Γ-X line appear quite different from the result of twinned samples.

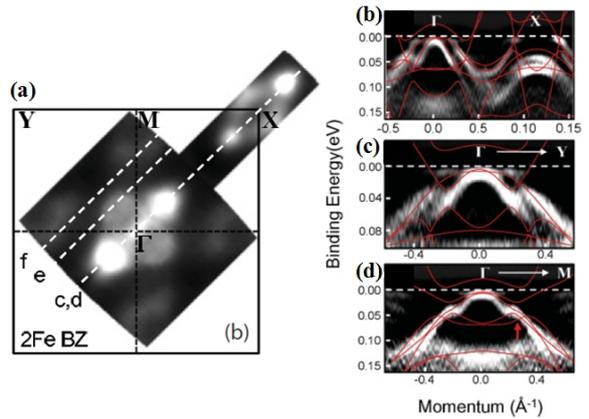

Fig. 14. Electronic structure of detwinned $BaFe_2As_2$ from ARPES and DFT. (a) Constant energy map at the Fermi level. (b)-(d) Band dispersions along high symmetry lines, overlain with calculated bands with Fe magnetic moment of 0.2 $\mu_B$. The As height was adjusted for the best fit and is larger by 0.070 Å$^{-1}$ than the experimental value. Calculated bands were renormalized by a factor of 3 and the Fermi level was shifted by 25 meV. Fig. reprinted from Ref. 86: Y. Kim *et al.*, Phys. Rev. B **83**, 064509 (2011). Copyright 2011 by the American Physical Society.

*Comparison of ARPES and DFT.*—Once the experimental dispersions are determined for detwinned antiferromagnetic samples, the results can be compared with the calculated spin-polarized band structure, producing indirectly the size of Fe magnetic moments. A comparison of experimental band dispersions from twinned sample and calculated band dispersions from LSDA+U with negative U suggested 0.5 $\mu_B$ for Fe magnetic moment in $Ba_{1-x}Sr_xFe_2As_2$ [87]. More recently, as shown in Figs. 14(b)-(d), ARPES band dispersions of detwinned samples are compared with calculated band dispersions using a constrained DFT method [86]. In this comparison, best match between experimental and calculated dispersions is achieved when the Fe magnetic moment is 0.2 $\mu_B$ for each Fe atom [86]

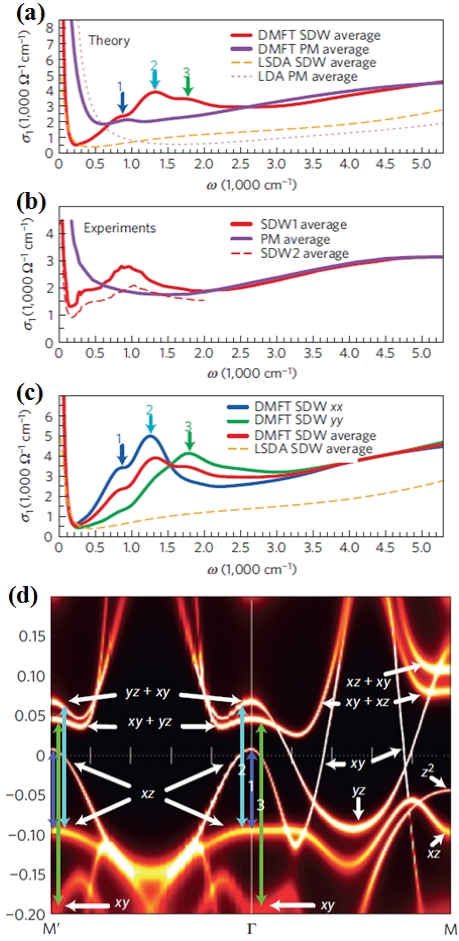

Fig. 15. Optical conductivity and ARPES of antiferromagnetic $BaFe_2As_2$ from DMFT calculation. (a) Calculated and (b) experimental in-plane average optical conductivity. (c) Calculated $xx$ and $yy$ optical conductivity. (d) Electronic spectra calculated by DFT+DMFT. Arrows mark the three types of optical transition that produce the three peaks in the optical conductivity. Fig. reprinted from Ref. 89: Z. P. Yin *et al.*, Nature Phys. **7**, 294 (2011). Copyright 2011 by Macmillan Publishers Ltd.

along with the arsenic height which is 0.07 Å larger than the experimentally measured value [88].

*DFT+DMFT calculations.*—The combination of density functional theory and dynamical mean-field theory (DFT+DMFT) [40] is used to study the electronic structures in $BaFe_2As_2$ in magnetic and non-magnetic phases and the results are compared with measured in-plane optical conductivity [89]. Fig. 15(a) shows the in-plane optical conductivity of $BaFe_2As_2$ in the spin-density-wave and paramagnetic phases calculated by both DFT+DMFT and standard DFT and then averaged over the two axes, *a* and *b*, to mimic twinned samples. Fig. 15(b) shows measured in-plane optical conductivity from Refs. 49 and 90. Both theory and experiments [49, 90] show a reduction of the low-frequency Drude peak with the emergence of magnetic ordering, which indicates a removal of a large fraction of carriers in the ordered state. DFT+DMFT calculation is successful in capturing qualitative features in experiments: a broad peak centered around 5,500 cm$^{-1}$ due to interband transitions, and additional peaks and shoulders below 2,000 cm$^{-1}$ at the spin-density-wave phase. These extra peaks and shoulders strongly depend on the polarization of light, as shown in Fig. 15(c), where the *x*-axis is the antiferromagnetic direction and the *y*-axis is the ferromagnetic direction [89].

## VII. Superconducting energy gap

An essential physical quantity of a superconductor is the superconducting energy gap, $2\Delta$, whose value and structure are intimately related to the pairing mechanism.

*Upper critical field.*—In the early stage, F. Hunte *et al.* reported resistance of $LaFeAsO_{0.89}F_{0.11}$ at high magnetic fields, up to 45 T, that show a remarkable enhancement of the upper critical field $B_{c2}$ compared with values expected from the slopes $dB_{c2}/dT \sim 2$ T/K near $T_c$, particularly at low temperatures [91]. Shown in Fig. 16(a) are the temperature dependences of the fields $B_{min}$, $B_{mid}$ and $B_{max}$ evaluated at 10 %, 50 % and 90 % of the normal state resistance at the transition temperature, respectively. F. Hunte *et al.* interpret that $B_{max}(T)$ and $B_{min}(T)$ reflect the temperature dependences of $B^{\parallel}_{c2}(T)$ and $B^{\perp}_{c2}(T)$, respectively, and $B^{\perp}_{c2}(T)$ exhibits a significant upward curvature, which is much less pronounced for $B^{\parallel}_{c2}(T)$. Since this behavior is similar to that observed in dirty MgB$_2$ films [92, 93], they suggested that superconductivity in LaFeAsO$_{0.89}$F$_{0.11}$ results from two bands: a nearly two-dimensional electron band with high in-plane diffusivity and a more isotropic heavy hole band with smaller diffusivity [91].

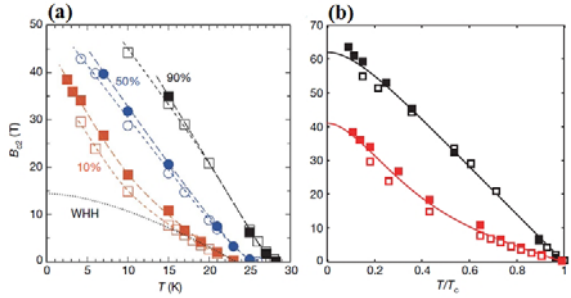

Fig. 16. Upper critical fields in LaFeAsO$_{0.89}$F$_{0.11}$. (a) The measured fields $B_{max}(T)$ and $B_{min}(T)$ (black and red squares, respectively) along with the midpoint transition fields (blue circles). The filled and open symbols correspond respectively to the parallel and perpendicular field orientations. The dotted line shows the WHH curve defined by the slope of $B_{min}(T)$ at $T_c$. (b) $B_{max}(T)$ (black squares) and $B_{min}(T)$ (red squares) vs $T/T_c$. Fig. reprinted from Ref. 91: F. Hunte *et al.*, Nature **453**, 903 (2008). Copyright 2008 by Macmillan Publishers Ltd.

*Andreev spectroscopy.*—The superconducting energy gap of SmFeAsO$_{0.85}$F$_{0.15}$ is measured by Andreev spectroscopy [94]. From the measurement, a single gap is observed in SmFeAsO$_{0.85}$F$_{0.15}$ with $T_c$ = 42 K. Fig. 17(a) shows result of an Au/SmFeAsO$_{0.85}$F$_{0.15}$ contact at 4.52 K within the bias voltage range of ±30 mV. There are two peaks with a separation of about 13.2 mV, indicating a single energy gap. The gap value of 2Δ = 13.34 ± 0.3 meV gives $2\Delta/k_BT_c$ = 3.68 (where $k_B$ is the Boltzmann constant), close to the Bardeen–Cooper–Schrieffer (BCS) prediction of 3.53. The gap decreases with temperature and vanishes at $T_c$ [Fig. 17(b)] in a manner consistent with the BCS prediction. The results indicate nodeless superconducting energy gap of a uniform size in different sections of the Fermi surface.

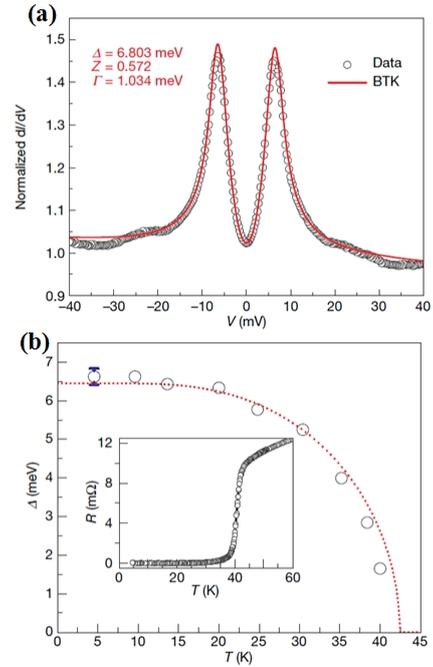

fig. 17. (a) Andreev spectra of Au/SmFeAsO$_{0.85}$F$_{0.15}$ contact at 4.52 K. Open circles are the experimental data, and the solid curve is the best-fit results. (b) Main panel, temperature dependence of the superconducting energy gap (open circles) in SmFeAsO$_{0.85}$F$_{0.15}$ obtained from the modified BTK fit at each temperature. The dotted curve is the BCS theory. Inset, the resistive transition at $T_c$. Fig. reprinted from Ref. 94: T. Y. Chen *et al.*, Nature **453**, 1224 (2008). Copyright 2008 by Macmillan Publishers Ltd.

*Superconducting energy gap in ARPES.*—ARPES can directly measure the $k$ dependence of superconducting energy gap. Up to now, ARPES measurements have been performed in high-quality single crystals of iron-based superconductors. H. Ding *et al.* have performed ARPES measurements on the superconducting Ba$_{0.6}$K$_{0.4}$Fe$_2$As$_2$ ($T_c$ = 37 K), and the results are shown in Fig. 18 [95]. Two superconducting energy gaps with different values were observed: a large gap (Δ ~ 12 meV) on the two small hole-type and electron-type Fermi-surface sheets, and a small gap (~ 6 meV) on the large hole-type Fermi surface. Both superconducting energy gaps are nearly isotropic without nodes, and close simultaneously at the bulk $T_c$. Since the large superconducting energy gap, which is ascribed to strong pairing interactions, is observed in the hole-type and electron-type Fermi surfaces connected by the spin-density-wave vector [95], H. Ding *et al.* suggest that the pairing mechanism originates from the inter-band interactions between these two nested Fermi-surface sheets. Similar experimental results have been reported by other groups [55, 96-101].

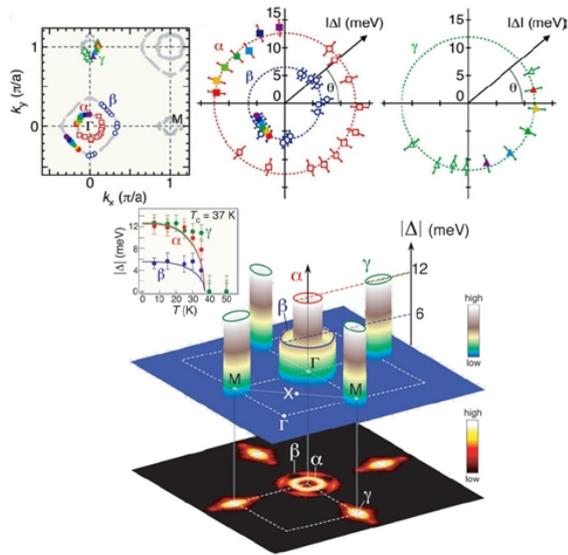

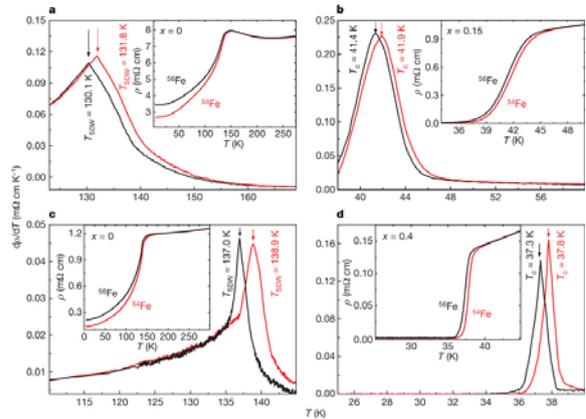

Fig. 18. (a) Fermi surface from ARPES and values of the superconducting energy gap ($\Delta$) in $Ba_{0.6}K_{0.4}Fe_2As_2$ at 15 K. (b) Three-dimensional plot of $\Delta$ measured at 15 K on the three observed Fermi surface sheets (shown at the bottom as an intensity plot) and their temperature evolutions (inset). Fig. reprinted from Ref. 95: H. Ding *et al.*, Europhys. Lett. **83**, 47001 (2008). Copyright 2008 by EDP Sciences.

## VIII. Pairing mechanism

*Too small electron-phonon interaction.*—For pure

Fig. 19. Resistivity, $\rho$, and its temperature derivative, $d\rho/dT$, of (a) SmFeAsO, (b) $SmFeAsO_{0.85}F_{0.15}$, (c) $BaFe_2As_2$, and (d) $Ba_{0.6}K_{0.4}Fe_2As_2$ isotopically substituted with $^{56}Fe$ and $^{54}Fe$. Obtained exponents are $\alpha = 0.37$ for magnetic transition temperature in SmFeAsO, $\alpha = 0.34$ for superconducting transition temperature in $SmFeAsO_{0.85}F_{0.15}$, $\alpha = 0.39$ for magnetic transition temperature in $BaFe_2As_2$, and $\alpha = 0.38$ for superconducting transition temperature in $Ba_{0.6}K_{0.4}Fe_2As_2$. The insets of (a) and (c) show remarkable difference in resistivity of $^{56}Fe$ and $^{54}Fe$ samples observed below the magnetic transition temperature, suggesting that iron isotope exchange has a strong effect on the magnetic state. Fig. reprinted from Ref. 103: R. H. Liu *et al.*, Nature **459**, 64 (2009). Copyright 2009 by Macmillan Publishers Ltd.

LaFeAsO, the calculated electron-phonon coupling constant $\lambda = 0.21$ and logarithmically averaged frequency $\omega_{ln} = 206$ K give a maximum $T_c$ of 0.8 K, using the standard Migdal-Eliashberg theory [102]. For the F-doped compounds, even smaller coupling constants are predicted. To reproduce the experimental $T_c$, a 5–6 times larger coupling constant would be needed. These results indicate that electron-phonon coupling is not sufficient to explain superconductivity in the whole family of iron-based superconductors.

*Isotope effects.*—As mentioned above, theoretical calculations indicate that the electron–phonon interaction is not strong enough to give rise to high $T_c$ in iron-based superconductors [102]. However, superconductivity and magnetism in iron-based superconductors show a strong sensitivity to the crystal lattice. R. H. Liu *et al.* reported effects of oxygen and iron isotope substitution on $T_c$ and magnetic transition temperature in SmFeAsO$_{1-x}$F$_x$ and Ba$_{1-x}$K$_x$Fe$_2$As$_2$ [103]. In these materials, the oxygen isotope effects on $T_c$ and magnetic transition temperatures are very small, while the iron isotope effects are substantial. Fig. 19 shows differences in the temperature dependence of resistivity, $\rho(T)$, and its derivative, $d\rho(T)/dT$, for SmFeAsO$_{1-x}$F$_x$ and Ba$_{1-x}$K$_x$Fe$_2$As$_2$ when $^{54}$Fe is substituted for $^{56}$Fe. Average values of the exponent for the magnetic transition temperature for several samples of SmFeAsO and BaFe$_2$As$_2$ are 0.39 and 0.36, respectively, and average values of the exponent for $T_c$ for several samples of SmFeAsO$_{0.85}$F$_{0.15}$ and Ba$_{0.6}$K$_{0.4}$Fe$_2$As$_2$ are 0.34 and 0.37, respectively. These results indicate that electron–phonon interaction plays some role in the superconducting mechanism by affecting the magnetic properties [104, 105].

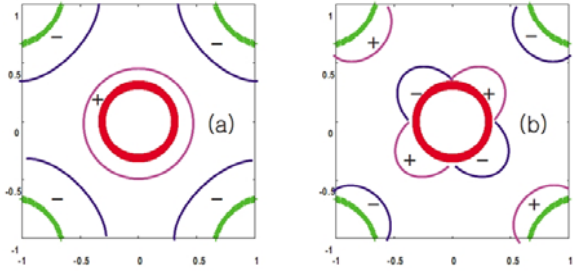

Fig. 20. ±*s*-wave gap and double *d*-wave gap. (a) ±*s*-wave gap and (b) double *d*-wave gap. Fig. reprinted from Ref. 113: Y. Bang and H.-Y. Choi, Phys. Rev. B **78**, 134523 (2008). Copyright 2008 by the American Physical Society.

*Theoretical models.*─To investigate possible pairing states, several theoretical models were proposed and many of them started with the orbital basis of the Fe 3*d* electrons including Hubbard interaction, *U*, and Hund coupling, *J* [106-112]. Some of these studies [106-108] found the ±*s*-wave gap [Fig. 20(a)] as a dominant instability. A *d*-wave gap [Fig. 20(b)] also often appears as a second instability [109, 110]. For comparison with experimental data, effects of different gap symmetries on the superconducting properties are extensively studied, including the density of states, temperature dependencies of spin-lattice relaxation rate $1/T_1$, Knight shift, and penetration depth [113]. Furthermore, Y. Bang *et al.* studied the impurity scattering on the ±*s*-wave superconductor and demonstrated that the ±*s*-wave gap state is the most consistent pairing state with available experimental data [114-116].

## IX. Phase of the superconducting energy gap

It is a crucial issue to measure the phase of the superconducting energy gap in iron-based superconductors since a strong candidate for the gap symmetry is ±*s* wave. Without phase sensitivity, the ±*s*-wave superconducting energy gap is indistinguishable from the conventional *s*-wave superconducting energy gap unless nodal planes exist on the Fermi surface. Indeed, penetration depth measurement [117] and ARPES [95, 118] suggest that the amplitude of the superconducting energy gap is finite all over the Fermi surfaces. Thus, the relative sign of the superconducting energy gap between the hole-type and the electron-type pockets of the Fermi surface should be determined by means of a phase-sensitive experiment in order to distinguish an unconventional ±s-wave from a conventional *s*-wave. So far, phase-sensitive experiments are very rare for iron-based superconductors.

*QPI under magnetic field.*─A phase-sensitive experiment, reported on a single crystal of FeSe$_{1-x}$Te$_x$, uses the scanning tunneling microscopy and spectroscopy to determine the sign of the superconducting energy gap by measuring the magnetic field dependence of the quasi-particle interference (QPI) patterns [119]. Because of the coherence factor C(**q**) representing the pairing of electrons, the scattering amplitudes are different for momentum **q** that connect the states with the same or opposite sign of the superconducting energy gap. If the scattering potential is even (odd) under time reversal, the QPI will appear only at momenta that connnect the states with the opposite (same) sign of the superconducting energy gap. In the ±s-wave theory, $q_2$ in Fig. 21(a) connects Fermi pockets with the opposite sign of the superconducting energy gap, whereas $q_3$ in the same figure connects Fermi pockets with the same sign of the gap. Therefore, in the ±s-wave theory, intensities of QPI at $q_2$ and $q_3$ should be suppressed and enhanced, respectively, with application of the magnetic field. The field-induced change in QPI intensities [Fig. 21(c)] shows opposite field dependences for $q_2$ and $q_3$, supporting the ±s-wave symmetry, that is, the sign of the superconducting energy gap is reversed between the hole-type and the electron-type Fermi-surface pockets.

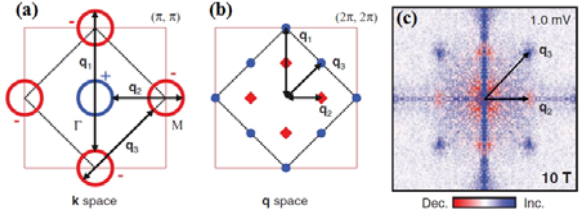
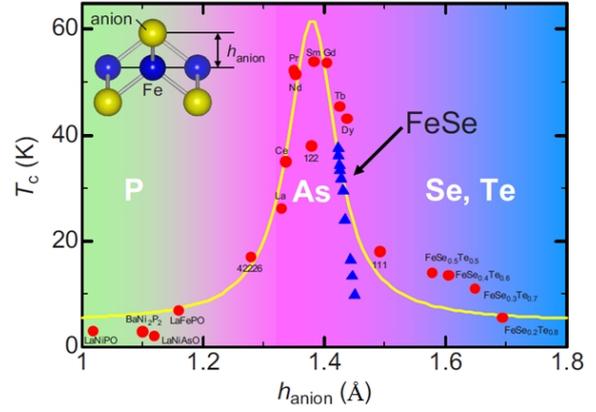

Fig. 21. Schematic of reciprocal-space electronic states of an iron-based superconductor and magnetic field-induced change in QPI intensities. (a) Fermi surface. There are two hole-type and two electron-type cylinders around Γ and M points, respectively. Signs of the superconducting energy gap expected for ±s-wave symmetry are shown by different colors. The arrows show inter-Fermi-pocket scatterings. (b) Expected QPI spots in **q** space due to inter-Fermi-pocket scatterings. Blue filled circles and red filled diamonds represent sign-preserving and sign-reversing scatterings, respectively, for the ±s-wave superconducting energy gap. (c) Magnetic field–induced change in QPI intensities with and without a magnetic field in $FeSe_{1-x}Te_x$. Fig. reprinted from Ref. 119: T. Hanaguri *et al.*, Science **328**, 474 (2010). Copyright 2010 by American Association for the Advancement of Science.

Fig. 22. $T_c$ vs anion height ($h_{anion}$) for various iron-based superconductors. Lanthanides (Ln) indicate LnFeAsO. 111, 122, and 42226 represent LiFeAs, $Ba_{0.6}K_{0.4}Fe_2As_2$, and $Sr_4Sc_2Fe_2P_2O_6$, respectively. The yellow line shows the fitting result. The inset shows a schematic view of $h_{anion}$. Fig. reprinted from Ref. 122: H. Okabe *et al.*: Phys. Rev. B **81**, 205119 (2010). Copyright 2010 by the American Physical Society.

## X. Optimal structure for superconductivity

In Fig. 22, plot of $T_c$ vs anion height ($h_{anion}$) for various iron-based superconductors [120-122] shows clear correlation between $T_c$ and $h_{anion}$, indicating the importance of anion positions in these iron-based superconductors. As the value of anion height increases, $T_c$ of the iron-based superconductors starts to increase dramatically up to ~55 K at a height of 1.38 Å, which corresponds to the optimum value of a 1111 system. However, beyond the optimum anion height (1.38 Å), $T_c$ decreases rapidly with increase of $h_{anion}$ [122]. Finally, the value of $h_{anion}$ becomes equal to that for non-superconducting FeTe (1.77 Å) [123]. It should be noted that superconductors with direct substitution in the $FeX_4$ tetrahedral layer or a large deviation from a divalent state ($Fe^{2+}$), e.g., an alkali-metal element or Co-doping samples of a 122 system [124] or chalcogen-substituted 11 system, are not particularly suitable for this trend [122]. From the correlation between $T_c$ and $h_{anion}$, it can be proposed that the electronic, magnetic, and superconducting properties of iron-based superconductors are inherently linked to their structural parameters, particularly the pnictogen height [122].

## XI. Summary

We briefly reviewed electronic, magnetic, and superconducting properties of iron-pnictide and iron-chalcogenide superconductors and their undoped parent compounds, covering only a small part of enormous works which have been achieved since 2008. Unlike high-$T_c$ copper-oxide superconductors, electronic structures and magnetic properties of iron-based superconductors are quite well understood. Since electron-phonon interaction is too weak to explain the superconducting transition temperature as high as 55 K, the iron-based superconductors provide an excellent chance to understand an unconventional superconductivity which might be helpful to reveal the origin of high-$T_c$ superconductivity in copper oxides.

**Acknowledgments**

This work was supported by NRF of Korea (Grant No. 2011-0018306).